\documentclass[12pt]{article}

\usepackage{url}

\usepackage{graphicx}
\usepackage{color}    
\usepackage{cite}
\usepackage{amsmath}
\usepackage{amssymb}
\usepackage{lineno,hyperref}
\modulolinenumbers[5]

\def\beq{\begin{equation}}
\def\eeq{\end{equation}}
\def\nn{\nonumber}
\def\bea{\begin{eqnarray}}
\def\eea{\end{eqnarray}}
\def\ba{\begin{array}}                  
\def\ea{\end{array}}

\newcommand{\vt}{\vert}
\newcommand{\lra}{\mu \leftrightarrow \tau}
\def\nn{\nonumber}

\evensidemargin \oddsidemargin
\begin{document}

\thispagestyle{empty}

\def\thefootnote{\fnsymbol{footnote}}


\vspace{1cm}

\begin{center}
{\large\sc {\bf The Breaking $\lra$ Symmetry through the ${\bf Q}_{6}$ Flavour Group}}

\vspace{1.4cm}

{\sc 
 J. C. G\'omez-Izquierdo$^{1,2,3}$\footnote{email: jcgizquierdo1979@gmail.com}, 
 F. Gonz\'alez-Canales$^{4}$\footnote{email: felixfcoglz@gmail.com} and  
 M~Mondragon$^{3}$\footnote{email: myriam@fisica.unam.mx}
}

\vspace*{1cm}

{\sl 
 $^{1}$~Departamento de Fisica y Matematicas, Tecnologico de Monterrey, Campus Estado de 
 Mexico, Atizapan de Zaragoza, Estado de Mexico, Apartado Postal 52926, Mexico.\\
 $^{2}$~Instituto~de Ciencias Nucleares, Universidad~Nacional Aut\'onoma de M\'exico, 
 M\'exico 3000, D.F., M\'exico.\\
 $^{3}$~Instituto~de F{\'{\i}}sica, Universidad~Nacional Aut\'onoma de M\'exico, 
 M\'exico 01000, D.F., M\'exico.\\
 $^{4}$~Departamento de F\'{\i}sica, Centro de Investigaci\'on y de Estudios Avanzados 
 del Instituto Polit\'ecnico Nacional, A.P. 14-740, 07000 M\'exico D.F., 
 M\'exico }

\end{center}

\vspace*{0.2cm}

\begin{abstract}
 In a supersymmetric scenario, we study masses and mixings for leptons through 
 the ${\bf Q}_{6}$ flavour symmetry. In the simplest case, the ${\bf M}_{\nu}$ 
 effective neutrino mass matrix, that comes from the type I see-saw mechanism, 
 breaks the $\mu \leftrightarrow \tau$ interchange symmetry. As consequence, the 
 reactor and atmospheric angles deviate from  $0^{\circ}$ and 
 $45^{\circ}$, respectively. At first glance, the model might accommodate very 
 well the reactor and atmospheric angles in good agreement with the experimental 
 data.
  
\end{abstract}

\def\thefootnote{\arabic{footnote}}
\setcounter{page}{0}
\setcounter{footnote}{0}
\newpage


\section{Introduction}
A new age in particle physics was opened with the discovery of neutrino 
oscillations and the precision measurements of corresponding parameters in 
this phenomena. Two of the main parameters that characterize the ordinary neutrino 
oscillations are the difference of the squared neutrino masses, as well as the 
flavour mixing angles. 
For the latter ones, the numerical values obtained as result of a global fit of 
current experimental data on neutrino oscillations~\cite{Forero:2014bxa}, at 
Best Fit Point (BFP) $\pm 1 \sigma$, are the following 

{\small
	\begin{equation}
	\begin{array}{l}\vspace{2mm}
	\sin^{2} \theta_{12} / 10^{-1} = 3.23 \pm 0.16, \quad
	\sin^{2} \theta_{23} / 10^{-1} = 
	\left\{ \begin{array}{l}\vspace{2mm}
	5.67_{-1.24}^{+0.32}   \\ 
	5.73_{-0.39}^{+0.25}  
	\end{array} \right. , \quad
	\sin^{2} \theta_{13} / 10^{-2}  = 
	\left\{ \begin{array}{l}\vspace{2mm} 
	2.26 \pm 0.12   \\  
	2.29 \pm 0.12  
	\end{array}  \right. ,
	\end{array}
	\end{equation}
} 
The values given in the upper (lower) row are for a normal
(inverted) hierarchy of the neutrino mass spectrum. This experimental
evidence was enough to show that neutrinos have a tiny mass, whereby
it was very easy to conclude that there is physics beyond the Standard
Model~(SM).

In the model building context, the $\lra$ flavour symmetry has been
widely used to 
propose possible extensions of the SM. In these extensions the $\lra$ flavour symmetry 
can be defined in two different ways: 
$i)$~the $\lra$ {\it permutation symmetry}~\cite{Mohapatra:1998ka} where the 
neutrino mass term is unchanged under the transformations 
$\nu_{e} \rightarrow \nu_{e}$, $\nu_{\mu} \rightarrow \nu_{\tau}$ and
$\nu_{\tau} \rightarrow \nu_{\mu}$. 
$ii)$~the $\lra$ {\it reflection symmetry}~\cite{Xing:2015fdg} where the 
neutrino mass term is unchanged under the transformations 
$\nu_{e} \rightarrow \nu_{e}^{c}$, $\nu_{\mu} \rightarrow \nu_{\tau}^{c}$ and
$\nu_{\tau} \rightarrow \nu_{\mu}^{c}$, where $c$ denotes the charge 
conjugation. 
But here we will only consider the first definition, hence in the
following 
when we mention the $\lra$ symmetry actually we mean  the $\lra$ permutation 
symmetry.

Historically,  theoretical physicists have proposed the $\lra$ symmetry  in 
order to reproduce the experimental data on lepton mixing angles. Namely, the 
$\lra$ symmetry is obtained if the reactor and atmospheric angles have the 
following values $\theta_{13} = 0^{\circ}$ and $\theta_{23} =45^{\circ}$, 
respectively.
%
However, this symmetry  is ruled out by the current 
experimental data, but these same data suggest some possibles
breakings of the 
$\lra$ symmetry which have been explored 
recently~\cite{Araki:2011zg,Gupta:2013it,Rivera-Agudelo:2015vza}. 
Besides, many discrete groups have been proposed in order to understand the 
underlying flavour symmetry behind the lepton mixing 
angles~\cite{Ishimori:2012zz}.

In this line of thought, we build a ${\bf Q}_{6}$ flavoured supersymmetric model 
to study masses and mixing for quarks and leptons where the $\lra$ symmetry is 
broken in the latter sector. As in early works on 
${\bf Q}_{6}$~\cite{Babu:2004tn}, it was necessary to extend the flavour label 
to the Higgs sector, this means that three families of doublets $H^{d}_{i}$ and 
$H^{u}_{i}$ are needed for the mixing. Our model is completely different from 
those already existing in the literature since the matter content assignment is 
very particular.
\section{The Model}
The matter content and their respective assignment under the 
${\bf Q_{6}}$ symmetry is displayed on Table~\ref{tabla1}.
\begin{table}[!htbp]
	\begin{center}
		\begin{tabular}{|c|c|c|c|c|c|c|} \hline \hline 
			{$\bf Q_{6}$}  & {\bf $1_{+,0}$} & {\bf $1_{+,2}$} & {\bf $1_{-,1}$} & 
			{\bf $1_{-,3}$} & {\bf $2_{2}$} & {\bf $2_{1}$} \\ \hline 
			Matter & {\footnotesize $H^{d}_{3}$ } & 
			{\footnotesize $H^{u}_{3}$ }, {\footnotesize $Y_{B}$ } & 
			{\footnotesize $L_{1}$ }, {\footnotesize $N^{c}_{1}$ }, 
			{\footnotesize $Q_{3}$ }, {\footnotesize $u^{c}_{3}$ } & 
			{\footnotesize $\ell^{c}_{1}$}, {\footnotesize $d^{c}_{3}$ }  & 
			{\footnotesize $L_{J}$}, {\footnotesize $\ell^{c}_{J}$ }, 
			{\footnotesize $N^{c}_{J}$ }, {\footnotesize $Q_{I}$ }, 
			{\footnotesize $d^{c}_{I}$ }, {\footnotesize $u^{c}_{I}$ }, 
			{\footnotesize $H^{d}_{I}$ } & {\footnotesize $H^{u}_{I}$ } \\ \hline\hline 
		\end{tabular}\caption{Matter Content.\label{tabla1}} 
	\end{center}
\end{table}
The present model is very peculiar in the sense that for the quarks and Higgs 
superfields, $Q_{I}$ and $H^{u,d}_{I}$ stand for a doublet under the flavour 
symmetry ${\bf Q}_{6}$ where $I=1,2$, this means explicitly for the former, 
$(Q_{1}, Q_{2})^{T}$. The rest of fields should be understood in the same way if 
they have the label $I$, otherwise, the fields are singlets under the flavour 
symmetry. On the other hand, for leptons, $L_{J}$ stands for a flavour doublet 
where $J=2,3$; and the first family belongs to any of the singlets. Then, the 
superpotential, which is allowed by the gauge and flavour symmetry, is given by 
{\small
	\begin{align}
	{\bf W} & 
	= y^{u}_{1} \left( Q_{1} u^{c}_{2} 
	- Q_{2} u^{c}_{1} \right) H^{u}_{3} 
	+ y^{u}_{2} \left( Q_{1} H^{u}_{2} 
	+ Q_{2} H^{u}_{1} \right) u^{c}_{3}
	+ y^{u}_{3} Q_{3} \left( u^{c}_{1} H^{u}_{2} 
	+ u^{c}_{2} H^{u}_{1} \right)
	+ y^{u}_{4} Q_{3} u^{c}_{3} H^{u}_{3} \nn \\
	&
	+ y^{d}_{1} \left[ Q_{1} 
	\left( - d^{c}_{1} H^{d}_{1} + d^{c}_{2} H^{d}_{2} \right) 
	+ Q_{2} \left( d^{c}_{1} H^{d}_{2} + d^{c}_{2} H^{d}_{1} \right) \right] 
	+ y^{d}_{2} \left( Q_{1} d^{c}_{1} + Q_{2} d^{c}_{2} \right) H^{d}_{3} 
	+ y^{d}_{3} Q_{3} d^{c}_{3} H^{d}_{3} \nn \\ 
	&
	+ y^{\ell}_{1} L_{1} e^{c}_{1} H^{d}_{3} 
	+ 	y^{\ell}_{2} \left[ L_{2} 
	\left( -e^{c}_{2} H^{d}_{1} + e^{c}_{3} H^{d}_{2} \right) 
	+ L_{3} \left( e^{c}_{2} H^{d}_{2} + e^{c}_{3} H^{d}_{1} \right) \right]
	+ y^{\ell}_{3} \left( L_{2} e^{c}_{2} + L_{3} e^{c}_{3} \right) H^{d}_{3} 
	+ y^{D}_{1} L_{1} N^{c}_{1} H^{u}_{3} \nn \\ 
	&
	+ y^{D}_{2} L_{1} \left( N^{c}_{2} H^{u}_{2} + N^{c}_{3} H^{u}_{1} \right) 
	+ y^{D}_{3} \left( L_{2} H^{u}_{2} + L_{3} H^{u}_{1} \right) N^{c}_{1}
	+ y^{D}_{4} \left( L_{2} N^{c}_{3} - L_{3} N^{c}_{2} \right) H^{u}_{3}
	+ y^{m} Y_{B} N^{c}_{1} N^{c}_{1} \nn \\ 
	&
	+ M_{R_{2}} \left( N^{c}_{2} N^{c}_{2} + N^{c}_{3} N^{c}_{3} \right) 
	\label{Yukb}
	\end{align}}
Comments are in order: one flavon $Y_{BK}$ (Babu-Kubo) has been included in 
order to build a flavour invariant Majorana mass matrix for the right-handed 
neutrinos (RHN's)~\cite{Babu:2004tn}; a subtle problem appears with our peculiar 
assignment for the matter content; there is no $\mu$ term for the Higgs sector 
since these are not flavour invariant, but these kind of terms should be present 
since they are crucial to get the electroweak symmetry breaking. In order to fix 
this, extra gauge singlets will be included to construct a gauge and flavour 
invariant $\mu$ term. Due to our interest in studying masses and mixings in this 
model, for the moment some alignments in the vacuum expectation values (vev's) 
of the scalars will be assumed. Then, we will leave aside the full scalar 
superpotential that eventually has to be analysed to have a complete model.
\section{Mases and mixings}
Going back to the expression in Eq.(\ref{Yukb}) we obtain the Dirac fermion mass 
matrices which have the following form 
{\small
	\begin{align}\label{M1}
	{\bf M}_{u}=
	\begin{pmatrix}
	0 & y^{u}_{1} \langle \textbf{H}^{u}_{3} \rangle & 
	y^{u}_{2} \langle \textbf{H}^{u}_{2} \rangle \\ 
	- y^{u}_{1} \langle \textbf{H}^{u}_{3} \rangle & 0 &
	y^{u}_{2} \langle \textbf{H}^{u}_{1} \rangle \\ 
	y^{u}_{3} \langle \textbf{H}^{u}_{2} \rangle & 
	y^{u}_{3} \langle \textbf{H}^{u}_{1} \rangle & 
	y^{u}_{4} \langle \textbf{H}^{u}_{3} \rangle
	\end{pmatrix}, \qquad 
	{\bf M}_{d} = 
	\begin{pmatrix}
	y^{d}_{2} \langle \textbf{H}^{d}_{3} \rangle 
	- y^{d}_{1} \langle \textbf{H}^{d}_{1} \rangle & 
	y^{d}_{1} \langle \textbf{H}^{d}_{2} \rangle & 0 \\ 
	y^{d}_{1} \langle \textbf{H}^{d}_{2} \rangle & 
	y^{d}_{2} \langle \textbf{H}^{d}_{3} \rangle 
	+ y^{d}_{1} \langle \textbf{H}^{d}_{1} \rangle & 0 \\ 
	0 & 0 & y^{d}_{3} \langle \textbf{H}^{d}_{3} \rangle
	\end{pmatrix},
	\end{align}}
and 
{\small
	\begin{align}\label{ME}
	{\bf M}_{D} = 
	\begin{pmatrix}
	y^{D}_{1} \langle \textbf{H}^{u}_{3} \rangle & 
	y^{D}_{2} \langle \textbf{H}^{u}_{2} \rangle & 
	y^{D}_{2} \langle \textbf{H}^{u}_{1} \rangle \\ 
	y^{D}_{3} \langle \textbf{H}^{u}_{2} \rangle & 0 & 
	y^{D}_{4} \langle \textbf{H}^{u}_{3} \rangle \\ 
	y^{f}_{3} \langle \textbf{H}^{u}_{1} \rangle & 
	- y^{D}_{4} \langle \textbf{H}^{u}_{3} \rangle & 0
	\end{pmatrix},\quad 
	{\bf M}_{\ell} = 
	\begin{pmatrix}
	y^{\ell}_{1} \langle \textbf{H}^{d}_{3} \rangle & 0 & 0 \\
	0 & y^{\ell}_{3} \langle \textbf{H}^{d}_{3} \rangle 
	- y^{\ell}_{2} \langle \textbf{H}^{d}_{1} \rangle & 
	y^{\ell}_{2} \langle \textbf{H}^{d}_{2} \rangle \\ 
	0 & y^{\ell}_{2} \langle \textbf{H}^{d}_{2} \rangle & 
	y^{\ell}_{3} \langle \textbf{H}^{d}_{3} \rangle 
	+ y^{\ell}_{2} \langle \textbf{H}^{d}_{1} \rangle 
	\end{pmatrix}.
	\end{align}}
In addition, the RHN mass matrix is diagonal in the flavour space, 
${\bf M}_{R} = \textrm{diag.} \left( M_{R_{1}}, M_{R_{2}}, M_{R_{2}} \right)$; 
here, $M_{R_{1}} = y^{n} \langle Y_{BK} \rangle$.

Now, assuming the degeneracy between 
$\langle \textbf{H}^{u}_{1} \rangle = \langle \textbf{H}^{u}_{2} \rangle$, and 
$\langle \textbf{H}^{d}_{2} \rangle = 0$, the resultant matrix, ${\bf M}_{u}$, 
has implicitly the NNI textures and the ${\bf M}_{d}$ mass matrix is diagonal. At the first glance the CKM mixing matrix will not be accommodated with great accuracy. Of course, we have to make sure that this is the case so that an $\chi^{2}$ analysis has to be done.
On the orden hand, the charged lepton mass matrix is also diagonal, then the PMNS 
mixing matrix comes only from the neutrino sector. Under these assumptions, we 
will focus on the lepton sector for the moment.
Therefore, 
{\small
	\begin{align}\label{ME2}
	{\bf M}_{D} = 
	\begin{pmatrix}
	a_{D} & b_{D} & b_{D} \\ 
	c_{D} & 0 & d_{D} \\ 
	c_{D} & -d_{D} & 0
	\end{pmatrix}, 
	\qquad 
	{\bf M}_{\ell} = 
	\begin{pmatrix}
	a_{\ell} & 0 & 0 \\
	0 & c_{\ell} - b_{\ell} & 0 \\ 
	0 & 0 & c_{\ell}+b_{\ell}
	\end{pmatrix} \equiv 
	\begin{pmatrix}
	a_{\ell} & 0 & 0 \\ 
	0 & B_{\ell} & 0\\ 
	0 & 0 & D_{\ell}
	\end{pmatrix}
	\end{align}
} 
and the effective neutrino mass matrix, that comes from the type I see-saw 
mechanism, is given by
{\small
	\begin{align}\label{mnu2}
	{\bf M}_{\nu} = 
	{\bf M}_{D} {\bf M}^{-1}_{R} {\bf M}^{T}_{D} = 
	\begin{pmatrix}
	m_{11} & m_{12} & m_{13} \\ 
	m_{12} & m_{22} & m_{23} \\ 
	m_{13} & m_{23} & m_{22}
	\end{pmatrix}, 
	\qquad 
	{\bf M}^{-1}_{R} \equiv \textrm{diag} (x, y, y).  
	\end{align}
}
Let us exhibit an appealing feature about ${\bf M}_{\nu}$: 
the block matrix $2-3$ provides a $\pi/4$ angle (to the mixing matrix) that may 
be identified with the atmospheric one. Moreover, if 
$\vt B_{\nu} \vt = \vt C_{\nu} \vt$ were true, ${\bf M}_{\nu}$ would have the 
$\mu-\tau$ symmetry, so that the PMNS mixing matrix would have 
$\theta_{13} = 0^{\circ}$ and $\theta_{23} = 45^{\circ}$ for the reactor and 
atmospheric angle, respectively.
Strictly speaking, in the present model, ${\bf M}_{\nu}$ does not possess the 
$\lra$ symmetry since $m_{12} \neq m_{13}$. This fact, actually, is crucial to 
get $\theta_{13} \neq 0^{\circ}$ and $\theta_{23} \neq 45^{\circ}$ in the PMNS 
matrix as we will see next.

To diagonalize ${\bf M}_{\nu}$, a perturbative analysis will be done as follows: 
${\bf M}_{\nu}$ is written as 
${\bf M}_{\nu} = {\bf M}^{0}_{\nu} + {\bf M}^{\delta}_{\nu}$ where
{\small
	\begin{align}\label{e4}
	{\bf M}^{0}_{\nu} =
	\begin{pmatrix}
	m_{11} & m_{12} & m_{12} \\ 
	m_{12} & m_{22} & m_{23} \\ 
	m_{12} & m_{23} & m_{22}
	\end{pmatrix} 
	\qquad \textrm{and} \qquad 
	{\bf M}^{\delta}_{\nu} = 
	\begin{pmatrix}
	0 & 0 & m_{13}-m_{12} \\ 
	0 & 0 & 0 \\ 
	m_{13}-m_{12} & 0 & 0
	\end{pmatrix}  
	\end{align}
}
The former mass matrix possesses the $\mu \leftrightarrow \tau$ symmetry. 
Here, we assume that ${\bf M}^{\delta}_{\nu}$ contains a perturbation parameter 
which will be defined later. In general, ${\bf M}_{\nu}$ is diagonalized by 
${\bf U}_{\nu} \approx {\bf U}^{0}_{\nu} {\bf U}^{\delta}_{\nu}$ with 
${\bf \hat{M}_{\nu}} = \textrm{diag}( m_{\nu_{1}}, m_{\nu_{2}}, m_{\nu_{3}} )
\approx {\bf U}^{\dagger}_{\nu} {\bf M}_{\nu} {\bf U}^{\ast}_{\nu}$, where 
${\bf U}^{0}_{\nu}$ diagonalizes to ${\bf M}^{0}_{\nu}$ and 
${\bf U}^{\delta}_{\nu}$ makes the same for the resultant matrix that depends on 
the difference $m_{12}\neq m_{13}$.

Explicitly, when the perturbation is switched off ($m_{12} = m_{13}$), 
we have
{\small
	\begin{align}\label{e5}
	{\bf U}^{0}_{\nu} = 
	\begin{pmatrix}
	\cos{\theta}_{\nu} e^{i \eta_{\nu} } & 
	\sin{\theta}_{\nu} e^{i \eta_{\nu} } & 0 \\ 
	- \frac{ \sin{\theta}_{\nu} }{ \sqrt{2} } & 
	\frac{ \cos{\theta}_{\nu} }{ \sqrt{2} } & 
	- \frac{1}{ \sqrt{2} } \\ 
	- \frac{ \sin{\theta}_{\nu} }{ \sqrt{2} }  & 
	\frac{ \cos{\theta}_{\nu} }{ \sqrt{2} } & 
	\frac{1}{ \sqrt{2} }
	\end{pmatrix} ~.
	\end{align} } 
At the same time, the matrix elements of ${\bf M}^{0}_{\nu}$ may be written in 
terms of neutrino mass eigenvalues and the $\theta_{\nu}$ angle as
{\small
	\begin{align}\label{e6}
	m_{11} & 
	= \left( m^{0}_{1} \cos^{2}{\theta}_{\nu} + m^{0}_{2} 
	\sin^{2}{\theta}_{\nu} \right) e^{2 i \eta_{\nu} }, \qquad 
	m_{12} = \frac{ \sin{\theta}_{\nu} \cos{\theta}_{\nu} 
		( m^{0}_{2} - m^{0}_{1} ) }{ \sqrt{2} } e^{i \eta_{\nu} }, \nonumber\\
	m_{22} + m_{23} & 
	= m^{0}_{1} \sin^{2}{\theta}_{\nu} + m^{0}_{2} \cos^{2}{\theta}_{\nu}, 
	\; \qquad 
	m_{22} - m_{23} = m^{0}_{3}.
	\end{align}
}
In general, the $m^{0}_{i}$ active neutrino masses are complex due to the 
presence of Majorana phases, and the $\theta_{\nu}$ angle is a free parameter. 
Going back to Eq. (\ref{e4}), in principle, $m_{12} \neq m_{13}$ so that it 
breaks the $\mu \leftrightarrow \tau$ symmetry but we will assume that 
$m_{13} - m_{12}$ is very small in order to apply a perturbative analysis.

Having pointed that, we rewrite ${\bf M}^{\delta}_{\nu}$ as
{\small
	\begin{align}\label{e7}
	{\bf M}^{\delta}_{\nu} = 
	\begin{pmatrix}
	0 & 0 & \sqrt{2}\bar{m}_{12}\delta \\ 
	0 & 0 & 0 \\ 
	\sqrt{2} \bar{m}_{12} \delta & 0 & 0
	\end{pmatrix},
	\quad \textrm{and} \quad  
	\delta \equiv \frac{ ( m_{13}-m_{12} ) / 2 }{ \bar{m}_{12} }  ,
	\end{align}}
where $\bar{m}_{12} = ( m_{13} + m_{12} )/2$ and $\vert \delta \vert\ll 1$, 
then, this will be our perturbation parameter. Thus, applying 
${\bf U}^{0}_{\nu}$ to ${\bf M}_{\nu}$ one obtains ${\bf U}^{0 \dagger}_{\nu}
{\bf M}^{0}_{\nu}{\bf U}^{0 \ast }_{\nu} + {\bf U}^{0 \dagger}_{\nu} 
{\bf M}^{\delta}_{\nu} {\bf U}^{0 \ast }_{\nu}$ where we can write explicitly as
{\small
	\begin{align}\label{e8}
	\textrm{diag}( m^{0}_{\nu_{1}}, m^{0}_{\nu_{2}}, m^{0}_{\nu_{3}} ) 
	+
	\begin{pmatrix}
	0 & 0 & \sqrt{2}\bar{m}_{12}  \cos{\theta}_{\nu} \delta e^{-i\eta_{\nu}} \\ 
	0 & 0 & \sqrt{2}\bar{m}_{12}  \sin{\theta}_{\nu} \delta e^{-i\eta_{\nu}} \\ 
	\sqrt{2} \bar{m}_{12} \cos{\theta}_{\nu} \delta e^{-i\eta_{\nu}} & 
	\sqrt{2} \bar{m}_{12} \sin{\theta}_{\nu} \delta e^{-i\eta_{\nu}} & 0
	\end{pmatrix} 
	\end{align}}
In this way, the active neutrino masses get corrections up to the second order 
in the $\delta$ parameter, and these are given by
{\small
	\begin{align}\label{e9}
	m_{\nu_{1}} = 
	m^{0}_{\nu_{1}} + \frac{2 \vert \bar{m}_{12} \vert^{2} \cos^{2}{\theta} 
		\vert \delta \vert^{2} }{ m^{0}_{\nu_{1}} - m^{0}_{\nu_{3}} }, \quad 
	m_{\nu_{2}} = 
	m^{0}_{\nu_{2}} + \frac{2 \vert \bar{m}_{12} \vert^{2} \sin^{2}{\theta} 
		\vert \delta \vert^{2} }{ m^{0}_{\nu_{2}} - m^{0}_{\nu_{3}} }, \nn \\
	m_{\nu_{3}} = 
	m^{0}_{\nu_{3}} + 2 \vert \bar{m}_{12} \vert^{2} \vert \delta \vert^{2} 
	\left[ \frac{ \cos^{2}{\theta} }{ m^{0}_{\nu_{3}} -m^{0}_{\nu_{1}} } 
	+ \frac{ \sin^{2}{\theta} }{ m^{0}_{\nu_{3} } - m^{0}_{\nu_{2}} } \right].
	\qquad
	\end{align}} 

At the same time, in Eq.~(\ref{e8}) the second mass matrix will modify the 
${\bf U}^{0}_{\nu}$ mixing matrix so that after a lengthy task the correction is 
written as
{\small
	\begin{align}\label{e10}
	{\bf U}^{\delta}_{\nu} = 
	\begin{pmatrix}
	N_{1} & N_{2} k_{1} k_{2} r_{2} \delta^{2}  & N_{3}k_{1}r_{1}\delta \\ 
	N_{1} k_{1} k_{2} r_{1} \delta^{2} & N_{2} & N_{3}k_{2}r_{2} \delta \\ 
	- N_{1} k_{1} r_{1} \delta & -N_{2} k_{2} r_{2} \delta  & N_{3}
	\end{pmatrix},
	\end{align}
}
where 
$r_{1,2} \equiv (m^{0}_{2}-m^{0}_{1})/(m^{0}_{3}-m^{0}_{1,2})$, 
$k_{1} \equiv \sin{\theta}_{\nu} \cos^{2}{\theta}_{\nu}$ and 
$k_{2} \equiv \sin^{2}{\theta}_{\nu} \cos{\theta}_{\nu}$. Besides, 
$N_{i}$ with $i=1,2,3$, stands for the normalization factor for each eigenvector 
of ${\bf U}^{\delta}_{\nu}$. At the end of the day, the full mixing matrix, 
which has to be compared with the standard parametrization, 
is ${\bf V}={\bf U}^{0}_{\nu}{\bf U}^{\delta}_{\nu}$.
%
Comparing the magnitude of entries of ${\bf V}$ with the mixing matrix in the 
standard parametrization of the PMNS,  give us the following expressions for 
the lepton mixing angles 
{\small
	\begin{align}\label{e12}
	\left| \sin \theta_{23}  \right|^{2}
	= \left| \frac{ N_{3} }{ \sqrt{2} } 
	\frac{ \left[ 1 - \cos{\theta}_{\nu} k_{2} r_{1} r_{2} \delta \right]}{ 
		\sqrt{ 1 - \vert \sin{\theta}_{13} \vert^{2} } } \right|^{2},  
	\qquad \qquad  
	\sin{\theta}_{12} 
	= N_{2} \frac{ \sin{\theta}_{\nu} }{ \sqrt{ 1 - 
			\vert \sin{\theta}_{13} \vert^{2} } }  
	, \nn \\
	\sin \theta_{13} 
	= N_{3} k_{1} r_{1} \cos{\theta}_{\nu} \delta  
	\left[ 1 + \tan{\theta}_{\nu} \frac{ k_{2} r_{2} }{ k_{1} r_{1} } 
	\right] .
	\qquad \qquad  \qquad  \qquad \qquad 
	\end{align}}
For simplicity, we neglected terms that are proportional to 
$\vert \delta \vert^{2}$ since we have assumed that $\vert \delta \vert\ll 1$. 
So far, these mixing angles depend directly of several free parameters namely: 
the active neutrino masses ($m^{0}_{i}$), the $\vert \delta\vert$ and 
$\theta_{\nu}$ parameters; and the Majorana and $\eta_{\nu}$ phases. Actually, 
the neutrino masses may be used as inputs in order to reduce the free 
parameters. At the end of the day, an $\chi^{2}$ analysis has to be done in 
order to explore the allowed regions for the free parameters.
\vspace{0.5cm}
\section{Conclusions}
We have constructed a supersymmetric model, with $Q_6$ flavour
symmetry and extended flavoured Higgs sector, where the breaking of the
$\mu-\tau$ symmetry leads to a deviation of $0^{\circ}$ and 
$45^{\circ}$ of the reactor and atmospheric angles respectively.  The
mixing angles depend on the active neutrino masses, as well  as on  the
difference of two of the neutrino mass matrix elements $\vert
\delta\vert$, the angle $\theta_{\nu}$, and the Majorana and
$\eta_{\nu}$ phases.  

The crucial difference with other discrete
symmetry models that use $Q_6$ or $S_3$ as symmetry group is that for
the quarks and Higgs fields we
have assigned the first two families to a doublet representation and
the third one to a singlet, but we have reversed the assignment for
the leptons, i.e. the second and third family are in a doublet and the
first in a singlet irrep.  This breaks the $\mu-\tau$ symmetry, thus
giving the possibility of realistic values for the reactor and atmospheric angles.
A $\chi^{2}$ analysis is in order to determine the experimentally
allowed regions in
parameter space. A complete model has to accommodate both lepton and quark sectors simultaneously, this work is still in progress.

\section*{Acknowledgements}
This work was partially supported by the Mexican grants  PAPIIT
IN111115 and Conacyt  132059. 
JCGI thanks Red de Altas Energ\'{\i}as-CONACYT for the financial support. 
FGC acknowledges the financial support from {\it CONACYT} under grant 23639.

\bibliographystyle{unsrt}
\bibliography{references.bib}
\end{document}